# Tensilely Strained Ge Films on Si Substrates Created by Physical Vapor Deposition of Solid Sources

Yize Stephanie Li[*] and John Nguyen

Department of Physics and Engineering, California State University, Bakersfield, CA 93311, USA

[*] Correspondence and requests for materials should be addressed to Y.S.L. (email: yli11@csub.edu)

**Abstract:**

The development of Si-compatible active photonic devices is a high priority in computer and modern electronics industry. Ge is compatible with Si and is a promising light emission material. Nearly all Ge-on-Si materials reported so far were grown using toxic precursor gases. Here we demonstrate the creation of Ge films on Si substrates through physical vapor deposition of toxin-free solid Ge sources. Structural characterization indicates that a high tensile strain is introduced in the Ge film during the deposition process. We attribute the presence of such a tensile strain to the difference in thermal expansion coefficient between Si and Ge. A Ge peak, centered at ~2100 nm, is evident in the photoluminescence spectra of these materials, which might result from direct band gap photoluminescence alone, or from superposition of direct band gap and indirect band gap photoluminescence. These Ge-on-Si materials are therefore promising in light emission applications.

**Introduction**

The development of Si-based electronics has reached a stage where the metal interconnect becomes a bottleneck hindering further advancement of computers and other modern electronics. Replacing electrical data transmission with optical data transmission is a promising route to tackle this problem and calls for the development of efficient Si-compatible lasers[1,2]. Group III-V semiconductors, owning to their direct band gap nature and well-established epitaxial thin film growth technique, have proven to be suitable materials for highly efficient lasers[3,4]. Nevertheless, as a result of the large lattice mismatch and incompatible thermal expansion coefficients between Group III-V materials and Si, integrating Group III-V lasers with Si-based electronics is a major technical challenge that has stimulated intense research efforts[5-12].

Ge, a Group IV semiconductor that is right below Si in the periodic table, is compatible with Si-based electronics. Ge, in its unstrained natural form, is an indirect band gap semiconductor. The indirect band gap (at the $L$ valley) energy is 0.664 eV, and the direct band gap (at the $\Gamma$ valley) energy is 0.800 eV, as shown in Fig. 1(a). Light emission is extremely inefficient in an indirect band gap semiconductor. To remove this roadblock in the development of Ge-based active photonic devices, a modification of the Ge band structure towards the transformation to a direct band gap semiconductor is essential. Tensile strain offers an effective route to achieving such a transition. With increasing tensile strain, the energy position of the $\Gamma$ valley decreases more rapidly than the $L$ valley [13-17]. With small biaxial tensile strain, such as 0.25% tensile strain, the difference between $\Gamma$ and $L$ valleys is less compared to that of unstrained Ge, although the material is still an indirect band gap semiconductor[14]. To achieve efficient light emission, the remaining difference needs to be compensated for by n-type doping[14]. On the other hand, with biaxial tensile strain of 2% or higher, theoretical calculations predict that the $\Gamma$ valley becomes even lower than the $L$ valley, making intrinsic Ge (i-Ge) a direct band gap material[15–17], as illustrated in Fig. 1(b). This scenario, dispensing with the need for degenerate doping which results in absorption losses of free carriers and nonradiative recombination, is desirable for lasing and light emission applications.

High tensile strain in Ge-on-Si materials has been demonstrated in suspended Ge microstructures fabricated by multi-step lithography processes[18,19]. These experiments served as a proof-of-concept for the role of high tensile strain in band structure modification in Ge materials. Nevertheless, the low-throughput fabrication process and/or the complex device structure made them difficult for practical applications. It would be favorable to induce high tensile strain directly in Ge films deposited on Si substrates, without going through any lithography process. However, only low tensile strain well below 1% was reported in these systems to date[20,21]. Moreover, the prevailing technique of Ge deposition on Si substrates requires using Ge precursor gases, such as $GeH_4$, in an ultra-high vacuum chemical vapor deposition (UHV-CVD) system. These gases are difficult to produce and/or to deliver and thus are expensive to purchase. They are also toxic and/or flammable.

In this work, we present a new approach of creating tensilely strained intrinsic Ge films on Si substrates through physical vapor deposition of solid Ge sources. Compared to the toxic Ge precursor gases used in the prevailing method, the solid Ge sources are toxin-free and environmentally friendly. Ge films created with this approach were epitaxial and possessed high crystalline quality. Biaxial tensile strain values between 2% and 3.3% were found to be present in these films, without any post-deposition treatment or lithography process. Such biaxial tensile strain could be explained by the difference in thermal expansion coefficient between Si and Ge. These Ge films exhibited direct band gap photoluminescence (PL), as expected for tensilely strained i-Ge films with a biaxial tensile strain value of 2% or higher, and are thus promising in lasing and light emission applications.

Results

**Epitaxial Ge films with high crystalline quality.** We created Ge films on native oxide-terminated or hydrofluoric (HF)- terminated Si(100) substrates, using solid Ge sources in a compact hot-wall chemical vapor deposition (CVD) system where Ar gas served as the carrier gas, as detailed in the Methods section. The thicknesses of our Ge films were typically between 300 nm and 600 nm. The as-grown Ge films were characterized by x-ray diffraction (XRD). Figure 2 shows a high resolution XRD 2θ-ω profile, a (004) reciprocal space map (RSM), and a (110) pole figure of a typical Ge film grown on a native oxide terminated- Si(100) substrate.

(400) peaks from the Si substrate and the Ge film are indicated in Fig. 2(a). The presence of a distinct film peak in the 2θ-ω profile confirms the formation of a Ge(100) film with high crystalline quality. Based on the 2θ value of the Ge(400) peak, the out-of-plane lattice constant of the Ge film is smaller than the natural lattice constant of Ge, indicating that in-plane tensile strain was induced in the Ge film. The strain values extracted from XRD 2θ-ω profiles were found to be consistent with those determined by Raman spectra (below). The well-defined separation of Ge and Si peaks in the RSM confirms that the shift of the Ge peak in Fig. 2(a) is due to tensile strain instead of intermixing of Ge and Si. The pole figure, which measures in-plane crystalline order, verifies that the Ge film is epitaxial, instead of polycrystalline.

Remarkably, XRD 2θ-ω profiles and other characterizations for samples grown on native oxide- terminated Si(100) substrates were found to be highly consistent with those on HF- terminated Si(100) substrates. It is therefore suggestive that, with the approach introduced in this work, HF treatment of Si substrates might not be necessary.

**Biaxial tensile strain above 2%.** The Raman spectra of the films, in the (001) backscattering configuration, were collected. In this configuration, the longitudinal optical phonon mode alone was detected[22]. Figure 3 shows normalized Raman spectra of two representative Ge films. Normalized Raman spectrum of a commercial bulk Ge wafer is also shown as a reference. The in-plane biaxial tensile strain values ($\varepsilon_\parallel$) in Fig. 3 were calculated from the Raman shift of the film ($\Delta\omega$) relative to the bulk Ge, using the equation $\Delta\omega = b\varepsilon_\parallel$, where b = -(415±40) cm$^{-1}$ [23]. A large in-plane biaxial tensile strain, above 2% (which is the calculated tensile strain for indirect band gap-to-direct band gap transition in Ge), was found to be present in a majority of our Ge films. As indicated above, similar strain values were also obtained from analysis of the XRD data on these films, substantiating the result of

Raman characterization. In-plane biaxial tensile strain up to 3.3% has been achieved in Ge films grown on both native oxide-terminated and HF- terminated Si(100) substrates. We note that, although intermixing of Ge and Si could result in Raman shift toward lower wavenumbers[24] and some degree of intermixing might occur during the deposition of these Ge films, intermixing should not be the dominant effect in our samples as is evident from the XRD data (discussed above) and the PL spectrum (shown below).

**Photoluminescence peak centered at ~2100 nm.** With biaxial tensile strain of 2% or higher, Ge is expected to transform to a direct band gap semiconductor. Figure 4 shows a room-temperature PL spectrum of a representative Ge film, with biaxial tensile strain of 2.52% as determined by its Raman spectrum. In the infrared regime, with the increase of wavelength ($\lambda$), the spontaneous emission rate decreases with $1/\lambda$ by cube power, and the competing nonradiative processes become more dominant [Capasso, F., private communication]. Moreover, infrared detectors targeted at wavelengths around 2000 nm or above are also less efficient, making PL detections challenging. The PL data in Fig. 4 is somewhat noisy, as expected for PL characterization in the wavelength regime of above 2000 nm using a PbS detector[25]. Modified Gaussian fitting and Lorentzian fitting, which fit the data reasonably well, are also shown in Fig. 4. A well-defined peak appears at ~ 2100 nm of the PL spectrum, corresponding to a band gap energy of ~ 0.59 eV. Band structure calculation in Ref. [16] suggested that, at the crossover between indirect and direct band gaps, the low temperature band gap energy is 0.53 eV. Our experimental result (at room temperature) thus shows a small deviation from the theoretical calculation at low temperature. The blue shift of our experimental data in comparison with the theoretical value could be a combination of effects from (1) occupation of higher-energy states in the $\Gamma$ valley due to limited density of states[19], (2) thermal energy (note that our PL measurement was performed at room temperature, instead of low temperature) induced carrier redistribution towards higher-energy states in the $\Gamma$ valley via Fermi-Dirac statistics[19], and (3) possible intermixing between Si and Ge at the substrate-film interface. The broad single peak shown in Fig. 4 might result from direct band gap PL alone, or from a superposition of direct band gap PL and indirect band gap PL that have similar wavelengths.

**Discussion**

With a 4.2% difference in natural lattice constants, the critical thickness of a Ge film grown on Si substrate is below 10 nm[26,27]. A Ge film is expected to be fully relaxed when several hundred nanometers of Ge are deposited on Si. The thermal expansion coefficient for Si is 2.6 x $10^{-6}$ °C$^{-1}$, and for Ge is 5.9 x $10^{-6}$ °C$^{-1}$. When the sample cools down naturally after the deposition, the substantial difference in thermal expansion coefficients between Ge and Si frustrates the shrinkage of Ge lattices, resulting in biaxial tensile strain in the Ge film. A similar effect has been reported for GaAs films grown on Si[28, 29]. We note that the biaxial tensile strain values found in our Ge films are 10 – 15 times larger than those reported in the literature[18, 21]. The solid Ge source, the hot-wall CVD system, and the relatively high pressure that we used for deposition are believed to account for the drastically higher tensile strain values in our samples. In comparison to the deposition of Ge films using precursor gases in a hot-wall system with much lower deposition pressures[21] or in a cold-wall system[18], the approach we developed in this work represents a highly dynamic process that enables creation of novel materials with unique structural and optical properties.

To summarize, we have created Ge films on native oxide- terminated and HF- terminated Si(100) substrates through physical vapor deposition of solid Ge sources. These Ge films exhibited high crystalline quality, biaxial tensile strain of 2% – 3.3%, and direct band gap photoluminescence. The high tensile strain in Ge films could be explained by the difference in thermal expansion coefficient between Si and Ge, with solid Ge serving as the source, in a hot-wall growth environment with relatively high deposition pressures. The structural properties and photoluminescence responses of the Ge films grown on native oxide- terminated substrates closely resembled those grown on HF- terminated substrates, suggesting that, with the approach that we developed, HF treatment of Si substrates before the deposition might not be necessary. The creation of epitaxial Ge films, with high tensile strain, on Si substrates using a toxin-free environmentally friendly approach, is important for the development of Si-compatible active photonic devices, a central enabling technology key to achieving optical data transmission.

**Methods**

**Film deposition.** A compact hot-wall chemical vapor deposition system was used to grow Ge films on native oxide- terminated or HF- terminated Si(100) substrates, using solid Ge sources (purity: 99.9999%). The Ge source and native oxide- terminated Si(100) substrates were cleaned with acetone, followed by isopropyl alcohol (IPA), in a ultrasonic water bath.

To prepare HF- terminated substrates, following acetone and IPA cleaning, the native oxide- terminated Si(100) substrates were dipped in a 49% HF solution for 30 seconds, and then rinsed in deionized water. The source and the substrates were placed in a quartz growth boat, which was then immediately loaded into the quartz growth tube of the CVD system. The growth tube was sealed and pumped down to $10^{-2} - 10^{-3}$ Torr using a mechanical pump. Ultra-high purity Ar gas (purity: 99.999%) was introduced into the growth tube to establish and maintain a pressure in the range of 0.5 – 10 Torr. The furnace temperature was increased from room temperature to 1000 °C and kept at 1000 °C for 30 – 60 minutes for the deposition of Ge films. The system was cooled down naturally when the growth was completed and the pressure in the growth tube was maintained the same until the temperature was below 200 °C. A schematic of our Ge deposition system and a representative atomic force microscopy (AFM) image of the Ge films are shown in Fig. 5.

**Structural characterizations.** We characterized the structural properties of as-grown Ge films by x-ray diffraction and Raman spectroscopy. A Panalytical MRD PRO Materials Research Diffractometer was used for XRD characterization. The Raman spectra of the films, in the (001) backscattering configuration, were collected by a LabRAM ARAMIS Raman Microscope, where only the longitudinal optical phonon mode was detected[22]. The wavelength of the excitation laser for Raman characterization was 633 nm.

**Photoluminescence measurements.** Room temperature PL measurements were carried out using a 532-nm continuous-wave excitation laser and a grating-based spectrometer equipped with a thermoelectric-cooled PbS detector. The pumping power was 500 mW, and the laser spot size on sample was ~ 1 mm, so the laser power density was ~ 250 W/cm$^2$.

**Acknowledgements**

This work is supported by the Research Council of the University (RCU) at California State University-Bakersfield. We thank Yiyin Zhou, Shui-Qing Yu, Youli Li, Elizabeth Powers, and Junhua Guo for experimental help.

**Contributions**

Y.S.L designed and carried out the experiment, analyzed the data, and prepared the manuscript. J.N. assisted with film deposition and XRD characterization.

**Competing Interests**

The authors declare no competing interests.

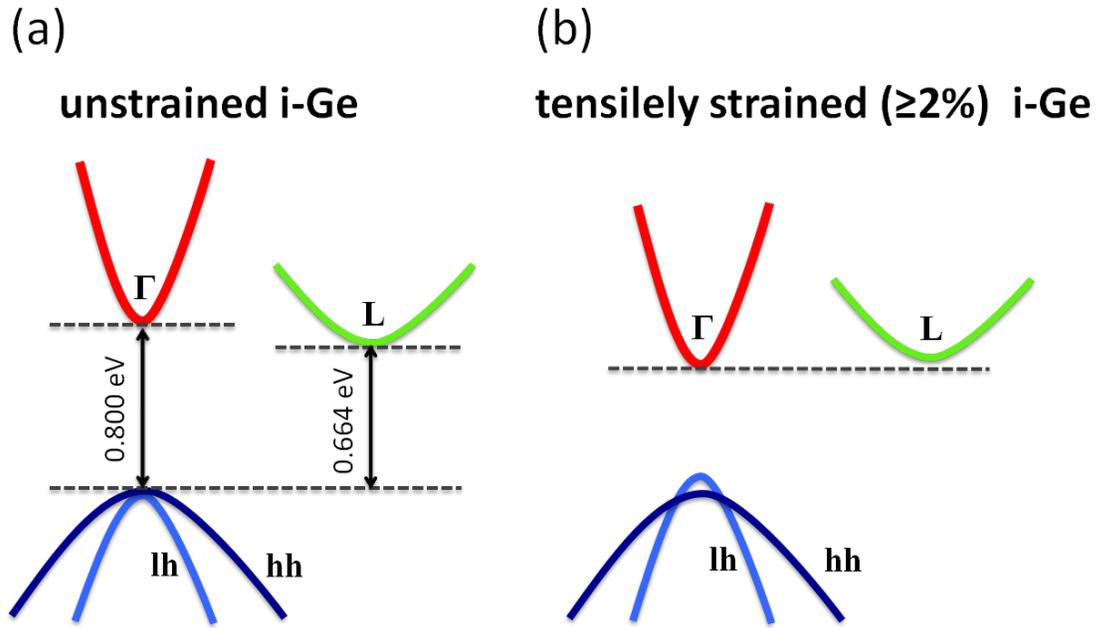

**Fig. 1** Schematic band structure of (a) unstrained intrinsic Ge (i-Ge) and (b) tensilely strained i-Ge with strain value ≥2%. The former is an indirect band gap semiconductor, while the latter is a direct band gap semiconductor.

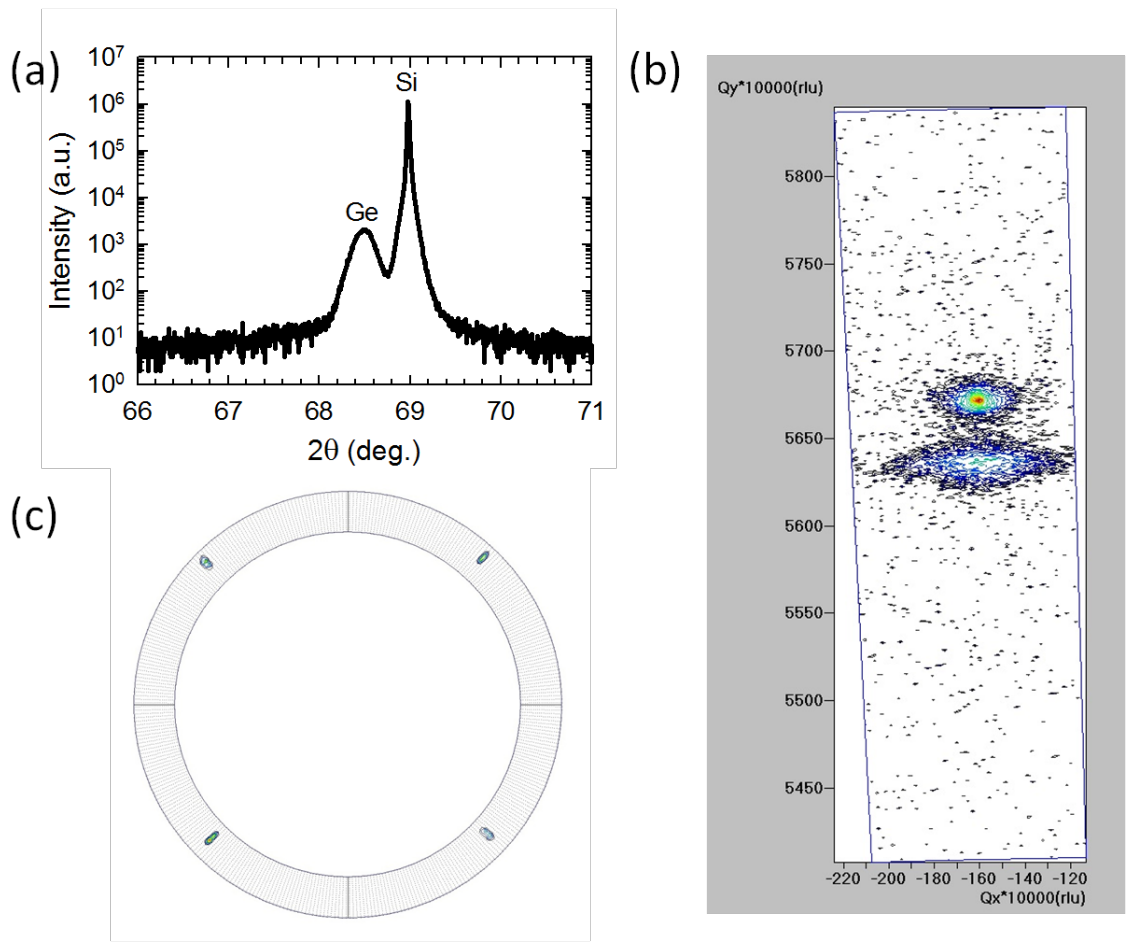

**Fig. 2** XRD characterization of a typical Ge film grown on a native oxide terminated- Si(100) substrate. (a) 2θ-ω profile of the sample. (b) (004) RSM of the same sample. (c) (110) pole figure of the same sample, where the 4 lines represent PHI rotation (from right) 0°, 90°, 180°, and 270°, and the inner and outer circles correspond to title angles PSI = 40° and 50°, respectively.

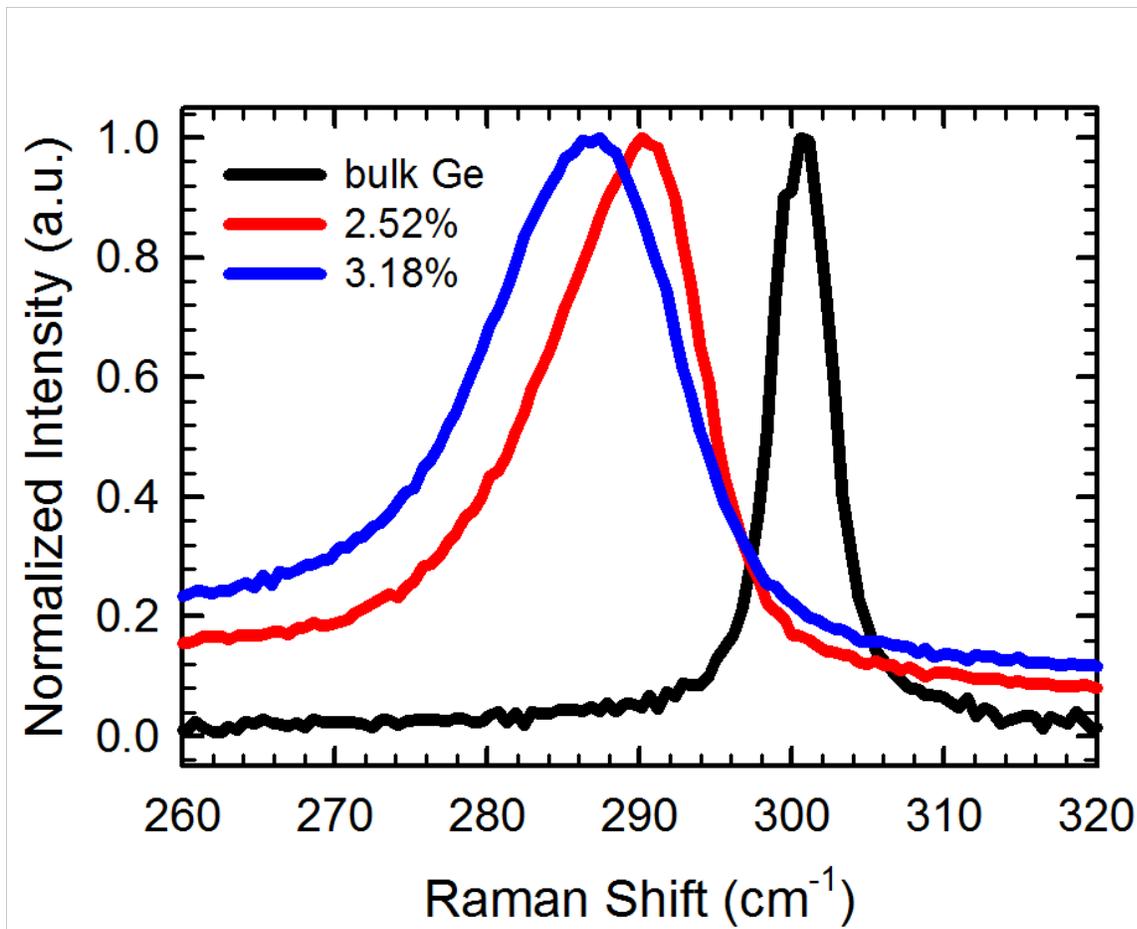

**Fig. 3** Normalized Raman spectra of tensilely strained Ge films grown on Si(100) substrates. Normalized Raman spectrum of bulk Ge is shown as a reference. The strain is calculated from the Raman shift of the film relative to bulk Ge.

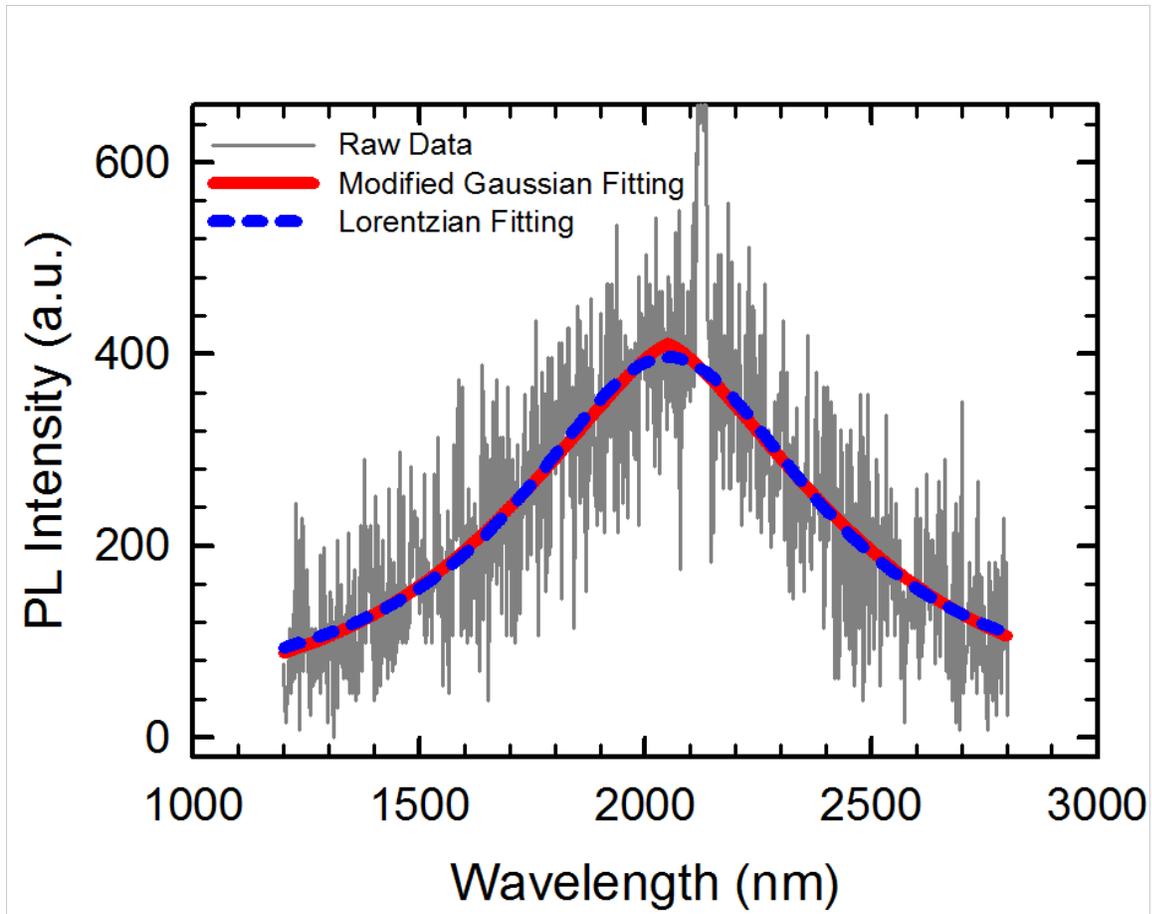

**Fig. 4** Room temperature photoluminescence spectrum of a Ge film with a tensile strain value of 2.52%, under the excitation of a 532 nm continuous wave pumping laser. The laser power was 500 mW, and the spot size on sample was about 1 mm. The spike at ~2128 nm came from fourth-order harmonic of the pumping laser. Modified Gaussian fitting and Lorentzian fitting of the experimental data are also shown.

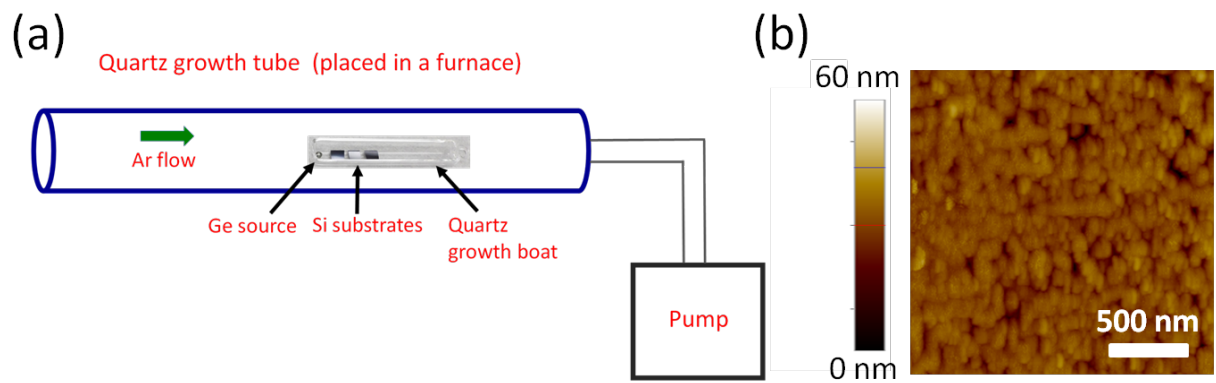

**Fig. 5** (a) Schematic of the Ge film deposition system. (b) A typical AFM image of a Ge film deposited on a Si substrate.